\documentclass[preprint,aps,draft,showpacs]{revtex4}

\usepackage{epsfig}

\begin{document}

\title{The fully self-consistent charge-exchange quasiparticle random phase 
approximation and its application to
the isobaric analog resonances}

\author{S. Fracasso and G. Col\`o}

\affiliation{Dipartimento di Fisica dell'Universit\`a degli Studi\\
             and INFN, Sezione di Milano, via Celoria 16\\
             20133 Milano, Italy}

\date{\today}

\begin{abstract}
A microscopic model aimed at the description of charge-exchange nuclear excitations
along isotopic chains which include open-shell systems, is developed. It consists
of quasiparticle random phase approximation (QRPA) made on top of 
Hartree-Fock-Bardeen-Cooper-Schrieffer (HF-BCS). The calculations are performed by using the Skyrme 
interaction in the particle-hole channel and a zero-range, density-dependent 
pairing force in the particle-particle channel. At variance with the (many) versions 
of QRPA which are available in literature, in our work special emphasis is put 
on the full self-consistency. Its importance, as well as the role played by the charge-breaking
terms of the nuclear Hamiltonian, like the Coulomb interaction, the 
charge symmetry and charge independence breaking (CSB-CIB) forces
and the electromagnetic spin-orbit, are elucidated by means of numerical 
calculations of the isobaric analog resonances (IAR). The theoretical energies of 
these states along the chain of the Sn isotopes agree
well with the experimental data in the stable isotopes. Predictions for
unstable systems are presented.
\pacs{24.30.Cz, 21.60.Jz, 21.30.Fe, 25.40.Kv, 21.10.Sf, 27.60.+j}

\end{abstract}

\maketitle

\section{INTRODUCTION}

The significant lack of knowledge concerning many properties of the charge-exchange 
nuclear excitations contrasts markedly with their importance for nuclear structure 
and the impact which they have on many interesting physical phenomena. 

The charge-exchange transitions involve a change in $N$ and $Z$ of the nucleus, 
keeping $A$ fixed. They can take place spontaneously, like in the well-known case of 
$\beta$-decay, or be induced by external fields when in a nuclear reaction a given 
amount of excitation energy $\Delta E$ and angular momentum $\Delta J$ is released to the 
nucleus. The spectra of charge-exchange reactions, like ($p$,$n$) or ($^{3}$He,$t$), are 
characterized by the emergence of collective isovector (i.e., $\Delta T$=1) giant 
resonances (IVGRs) in analogy with the non charge-exchange case~\cite{Harakeh}. 
However, a unifying picture of these $\Delta T_z$=$\pm$1 states is still, to a large 
extent, missing. For instance, the $\Delta L$=0 charge-exchange isovector giant monopole  
resonance (IVGMR) is one of the most elusive nuclear states, despite a long series 
of experiments aimed at its identification~\cite{IVGMR}; at the same time, its 
knowledge would be important for the determination of the ground state isospin 
mixing. Also the higher multipoles, that is, the charge-exchange dipole, quadrupole 
and octupole resonances, are basically unknown. This is mainly due to the lack of 
really selective probes: in particular, the separation of the electric (i.e., 
$\Delta S$=0 or ``non spin-flip'') and magnetic (i.e., $\Delta S$=1 or ``spin-flip'') modes
is far from being trivial. On the other hand, a systematic pattern of the energy and
collectivity of these states would shed light on the strong uncertainties 
concerning the isovector part of the nucleon-nucleon ({\itshape NN}) effective interaction and
the symmetry part of the nuclear equation of state.

It has to be mentioned that knowing the properties of the nuclear charge-exchange
states allows also to attack other kinds of problems outside the realm of nuclear
structure. These states enter the description of the double $\beta$-decay, and 
the need of a reliable theory of this process is a longstanding problem. More
generally, all the weak interaction processes within atomic nuclei involve 
charge-exchange transitions as far as charged currents are involved. We have in 
mind many processes which are of interest for neutrino physics, like the interaction
of these peculiar particles with nuclei, or for astrophysics, like the mechanisms 
which are responsible for the evolution of neutron stars or the $\beta$-decay of isotopes which lie on
the $r$-process path of stellar nucleosynthesis. 

A significant exception to the unsatisfactory ignorance of the charge-exchange IVGRs  
is provided by the availability of many experimental data on the isobaric analog 
resonance (IAR) and the Gamow-Teller resonance (GTR). The IAR is the simplest 
charge-exchange transition, in which a neutron is changed into a proton without 
any other variation of the quantum numbers (that is, $\Delta J$=$\Delta L$=$\Delta S$=0). 
The corresponding operator is
\begin{equation}
{\hat O}_{{\mbox{\upshape IAR}}} \equiv \sum_{i=1}^A t_-(i),
\label{op_iar}
\end{equation}
namely it is the usual Fermi, or isospin-lowering, operator. In the Gamow-Teller case, 
the transition is accompanied by a spin-flip ($\Delta L$=0, $\Delta J$=$\Delta S$=1), 
and the operator is
\begin{equation}
{\hat O}_{{\mbox {\upshape GTR}}} \equiv \sum_{i=1}^A \vec\sigma t_-(i).
\end{equation}
Many data coming at an early stage from ($p$,$n$), and later from other reactions, have 
shown that these resonances can be systematically identified in the isotopes with neutron 
excess (in which the corresponding $t_+$ transitions are Pauli-blocked). The IAR 
consists of a single, very narrow peak, whereas the GTR manifests itself with a broad bump
and can also be fragmented in different peaks. Experimentally, when the incident 
projectile energy is increased, the excitation of the GTR is favoured over the IAR; this
experimental fact has allowed to establish that the strengths of the spin-independent and 
the spin-dependent 
components of the effective {\itshape NN} interaction have different 
behaviour as a function of
the energy. 

From this rather general introduction, the motivation for microscopic calculations 
of the charge-exchange states in nuclei is already evident. We must 
add that one of the main present interests in 
nuclear physics is the understanding of the limits of nuclear stability, and of the 
exotic, very neutron-rich (or proton-rich) nuclei, that is, of the systems with different values 
of $N-Z$ than those which characterize the valley of stability. 
The experimental evidences about the isospin properties
of exotic nuclei are still rather scarce. In order to make predictions in this delicate sector
accurate calculations are called for, which do not make approximations by neglecting
terms of the nuclear Hamiltonian in an uncontrolled way. 

For nuclei with mass up to $A\sim$50, the shell model (SM) calculations can be rather 
successful and are indeed performed, also in the cases of interest for 
applications. The agreement with the experimental findings (like the GT strength
and/or the $\beta$-decay half-life) can be quite good~\cite{sm}. 
However, these calculations become too demanding, or impossible, for heavier nuclei. 
Also, they have trouble if the space must be large enough so to account for high-energy
transitions; these transitions can be induced, for instance, by neutrinos which
follow a supernova explosion. In Ref.~\cite{volpe1} it has been shown that for energies
above 50 MeV the SM calculations may underestimate the strength of the charge-exchange 
transitions. 

The alternative choice is a mean-field based calculation which employs an effective 
{\itshape NN} interaction. In this case, the ground state of the parent ($N$,$Z$) system is obtained 
by means of a Hartree-Fock (HF) calculation, extended to Hartree-Fock-Bardeen-Cooper-Schrieffer 
(HF-BCS) or Hartree-Fock-Bogoliubov (HFB) in the case of open-shell nuclei where pairing 
is relevant. In the two cases, respectively, the charge-exchange excited states 
in the $(N\mp 1, Z\pm 1)$ isobars can be obtained within the framework of the linear 
response theory, that is, by using the random phase approximation (RPA) 
or its extension to the pairing case, namely the quasiparticle RPA (QRPA). These are 
well-known theories, whose general features can be found in many textbooks. However, 
there are only few examples, if any, of fully self-consistent QRPA calculations --- which 
constitute the proper scheme for the analysis of long isotopic chains extending towards 
the drip lines. In fact, self-consistency is a crucial issue if the calculations are 
required to have predictive power far from the experimentally known regions of the mass 
table. Moreover, as we discuss below, self-consistency plays a special role if the 
isospin symmetry and its breaking enters the discussion. We repeat here that 
self-consistency means that the residual particle-hole ({\itshape p-h}) and particle-particle ({\itshape p-p}) 
residual forces, which enter the QRPA equations (cf. Sec. II), are derived from the same
energy functional from which the HF-BCS of HFB equations describing the ground state
are obtained. 

The first attempt of self-consistent QRPA on top of HFB is found in Ref.~\cite{engel99}. 
The Skyrme zero-range force and a zero-range pairing interaction are used, respectively,
in the mean-field and in the pairing channel to solve the HFB equations in coordinate
space (cf. also Ref.~\cite{hfb_doba}). The associated QRPA equations are solved in the canonical
basis. The method is applied to the calculation of Gamow-Teller $\beta$-decay half-lives.
These 1$^+$ states are known to be sensitive only to the $T$=0 component of the residual {\itshape p-p} 
interaction, if pairing is described by means of a zero-range force. In 
Ref.~\cite{engel99} it is assumed that, since this $T$=0 pairing does not manifest itself 
in the HFB ground state of nuclei with $N$ different from $Z$ by a few units, one is
free to introduce it within QRPA in a completely different way than the $T$=1 pairing, 
without any constraint related to self-consistency. The authors have employed a finite-range
interaction with free parameters: the overall strength is fitted to reproduce some selected
$\beta$-decay experimental findings. The same approach is used in Ref.~\cite{bender02} 
to analyze the performance of existing Skyrme parametrizations in the case of the GT resonances, 
and to correlate it with their ability to reproduce the values of 
the empirical Landau parameters of infinite matter.

In the present paper, we would like to discuss the implementation of a fully self-consistent 
charge-exchange QRPA by putting emphasis on aspects which were not considered in 
Refs.~\cite{engel99,bender02}. A first aspect is the issue of isospin invariance. We
show that the $T$=1 component of the residual {\itshape p-p} force can be fixed by exploiting
this invariance. Our hypothesis is supported by the absence of strong evidences coming from 
literature which point to a clear
 need to differentiate the strengths of the 
three components of the $T$=1 pairing. Within this assumption, we show that we can
obtain results for the IAR which are quite satisfactory when compared with experiment. 
The IAR is a serious benchmark for every theoretical model, because of its intimate 
relationship with the isospin symmetry (cf., e.g., Ref.~\cite{auerbach_report}). 
In fact, if the whole Hamiltonian $H$ commuted with 
isospin, and if one were able to solve $H$ exactly, the resulting IAR would be degenerate 
with the parent ground state. Many of the approximation schemes which are commonly used to 
solve the nuclear many-body problem destroy this property of the Hamiltonian. 
HF and HF-BCS belong to this category and introduce a spurious isospin breaking 
(as soon as $N\neq Z$ in case of HF). 
Instead, it has been demonstrated that self-consistent RPA and QRPA calculations 
restore the isospin symmetry and eliminate any spuriousity~\cite{iso_restore}, being 
in this sense ``good'' symmetry-preserving approximations. Consequently, only 
within their framework it is possible to assess the relative importance of 
the physical contributions which are responsible for an explicit isospin breaking: the Coulomb force, 
the electromagnetic spin-orbit, and the other 
charge-symmetry breaking (CSB) and charge-independence breaking (CIB) terms in the 
nuclear Hamiltonian. The study of these issues in the case of the IAR for the 
open-shell isotopes 
is an original feature of the present work. 
Since we do not go beyond QRPA, we cannot discuss
the (narrow) width of the IAR. The extensions of RPA and of the normal, non charge-exchange 
QRPA, intended to take into account the coupling with more complex configuration and 
therefore to describe the spreading width of the resonances, are described 
elsewhere (see the references quoted in Sec. II). 

In our work, we employ zero-range forces. We are not aware of self-consistent calculations
of charge-exchange states done by using finite-range interactions like Gogny. On the other
hand, in recent papers the relativistic mean-field (RMF) effective Lagrangians, based on
the description of nucleons as Dirac particles which interact by means of the exchange 
of effective mesons, have been used for the calculation of the IAR and the 
GTR~\cite{paar0304}, as well as of $\beta$-decay rates~\cite{niksic05}. The RMF 
description of the ground state and of the excited nuclear states emerges from 
rather different ingredients than those which characterize the non-relativistic 
mean-field. It is known that the isovector channel of the {\itshape NN} interaction, and
consequently the symmetry part of the energy functional, are quantitavely 
not the same, generally speaking, in the two cases (the symmetry energy at saturation
and its derivatives are generally larger in the relativistic case). In the relativistic
calculations of the spin and isospin excitations the pion-exchange is very important; but
this degree of freedom is not present in the ground state description because of
parity conservation. On the other hand, in the case of RMF the spin-orbit is 
automatically considered, at variance with the non-relativistic case. Finally, we 
are not aware of attempts to include CSB and CIB forces in the RMF calculations. 
All this should be kept in mind when comparing our results with those of~\cite{paar0304}. 

\section{THEORETICAL FRAMEWORK}

As mentioned in the previous Section, charge-exchange RPA and QRPA are well-known and described
in textbooks. We try here to recall only the basic elements, or some details which are 
useful for the following discussion. 

In the case of charge-exchange RPA, self-consistent calculations have been available 
for many years. In particular, the first application to the case of the IAR can be found in
Ref.~\cite{auerbach_proc}. Extensive calculations of the response to different 
multipole operators, made by using the coordinate
space formulation of RPA with proper treatment of the particle continuum, but dropping
for simplicity some terms of the residual interaction, are reported in~\cite{auerbach}. 
As we have recalled in the Introduction, 
it is well-known that mean-field calculations of this kind cannot reproduce the total width
of the resonances, but only the escape width if the continuum is correctly taken into
account. The spreading width, associated with the coupling of the simple {\itshape p-h} configurations
to the more complex states, of two particle-two hole (2$p$-2$h$) character, can be described
only by diagonalizing the effective Hamiltonian in a larger model space than the one of
RPA. A microscopic model suited for this purpose has been developed in~\cite{adachi,colo94}. 
In~\cite{colo98} the importance of CSB and CIB forces for the IAR width has been studied. 

In the case of the QRPA, most of the charge-exchange calculations performed so far make use 
of two separable {\itshape p-h} and {\itshape p-p} residual interactions (having, as a rule, the same functional
form and two different overall parameters $g_{ph}$ and $g_{pp}$), 
as in the pioneering work by J.~A. Halbleib and 
R.~A. Sorensen~\cite{halbleib}, where the formalism has been developed for the first time. 

We start by solving the HF-BCS equations in coordinate space by using a radial mesh 
extending up to 20 fm (with a step of 0.1 fm). The
HF equations contain the Skyrme {\itshape NN} interaction and we have chosen in this work the 
parametrization SLy4~\cite{chabanat}, which has been determined by trying to retain many of the
advantageous features of the previous versions of the Skyrme force, as well as by fitting 
the equation of state of pure neutron matter obtained by means of realistic
forces. This latter characteristic should justify its use for systems outside
the valley of stability. The BCS equations are solved, as usual, in a limited space:
only the levels which correspond to the 82--126 neutron shell are included. 
The pairing force that we have used is of the type
\begin{equation}
V=V_0 \left( 1- \left( {\varrho \left( {\vec r_1+\vec r_2\over 2} \right)
\over \varrho_c} \right)^\gamma \right)\cdot\delta(\vec r_1-\vec r_2).
\label{ppforce}
\end{equation}
The parameter $\gamma$ is fixed to one for the sake of simplicity. With the same spirit,
$\varrho_c$ is set at 0.16 fm$^{-3}$. The strength $V_0$ has been determined by requiring
a reasonably good agreement between the theoretical and empirical values of the pairing
gaps $\Delta$ along the whole series of isotopes under study. This agreement, when $V_0$ is
equal to our adopted value of 680 MeV~fm$^{3}$, is shown in Fig.\ \ref{average_delta}. 
We notice in this context that a rather similar pairing force, having 
$V_0$=625 MeV~fm$^{3}$, has been used independently 
by other groups to carry out large-scale, systematic calculations of the pairing gaps
and of the rotational bands (see~\cite{duguet} and references therein). 
It is known that the HFB treatment is more coherent than the HF-BCS one; however, qualitatively
important differences between the results of the two methods show up only in the case of
weakly bound nuclei, which will not be considered in the present study. 

When the ground state is obtained, together with the filled or partially occupied 
states lying within the pairing window, a number of unoccupied states (which have occupation
factors $v^2$ strictly equal to zero) are calculated by using spherical box boundary conditions.
This means that our continuum is discretized. For
every value of $(l,j)$, we calculate unoccupied states with six increasing values of $n$. 
The dimension of the space has been checked by looking at the results for the energy and
the strength of the IAR, which have been found to be stable when we enlarge the space, 
by considering in some cases up to ten increasing values of $n$. We have checked that 
also the $N-Z$ sum rule is accurately reproduced. In this configuration space, the QRPA matrix equation 
written on the basis made up with the two quasiparticle states having good angular momentum
and parity $J^\pi$, reads 
\begin{equation}
\left( \begin{array}{cc} A & B \\ -B & -A \end{array} \right) 
\left( \begin{array}{c} X^{(n)} \\ Y^{(n)} \end{array} \right) =  E_n 
\left( \begin{array}{c} X^{(n)} \\ Y^{(n)} \end{array} \right). 
\end{equation}
In this formula, $E_n$ is the energy of the $n$-th QRPA state in the parent nucleus, 
while $X^{(n)}$, $Y^{(n)}$ are the corresponding forward- and backward-amplitudes. 
The matrices $A$ and $B$, in the angular momentum coupled representation, have the explicit form
\begin{eqnarray}
A_{pn,p'n'} & = & (E_p+E_n)\delta_{pp'}\delta_{nn'} + \nonumber \\
            &   & + V^{(J)}_{pnp'n'} (u_{p}u_{n}u_{p'}u_{n'}+v_{p}v_{n}v_{p'}v_{n'}) + \nonumber \\
            &   & + W^{(J)}_{pnp'n'} (u_{p}v_{n}u_{p'}v_{n'}+v_{p}u_{n}v_{p'}u_{n'}), \nonumber \\
B_{pn,p'n'} & = & - V^{(J)}_{pnp'n'} (u_{p}u_{n}v_{p'}v_{n'}+v_{p}v_{n}u_{p'}u_{n'}) +  \nonumber \\
            &   & + W^{(J)}_{pnp'n'} (u_{p}v_{n}u_{p'}v_{n'}+v_{p}u_{n}v_{p'}u_{n'}).
\end{eqnarray}
Here, the indices $p$ and $p'$ ($n$ and $n'$) refer to proton (neutron) quasiparticles.
$E$ is their energy, whereas $u$ and $v$ are the usual BCS occupation factors. $V^{(J)}$ and
$W^{(J)}$ indicate respectively the coupled 
{\itshape p-p} and {\itshape p-h} matrix elements. The {\itshape p-h} matrix elements are derived from
the Skyrme part of the energy functional: all the terms are considered, including the two-body 
spin-orbit. 

The {\itshape p-p} matrix elements, when consistently derived from the energy functional, are 
those of the bare force (\ref{ppforce}): in fact, no rearrangement terms show up if we do
not impose any dependence on the anomalous density in the force itself. The isospin invariance
that we have assumed, 
demands that the $T$=1 component of the pairing force is the same in the three channels: 
neutron-neutron, proton-neutron and proton-proton. In the present case, since we
deal with the Sn isotopes which have magic proton number, there is no proton pairing in the
ground state. Also, we have neglected proton-neutron 
pairing in the ground state: in fact, this may be important
only in nuclei having $N\sim Z$ and we 
have considered Sn isotopes in the
range 104$\le A \le$132. However, the proton-neutron $T$=1 pairing force enters the
QRPA equations (in the $V$ matrix elements) and we can say that we have
preserved the self-consistency in the pairing channel, in the same way as in the {\itshape p-h}
one. 

The CSB and CIB forces are included in our HF-BCS iterative procedure. These forces are
parametrized as in Ref.~\cite{sagawa}, where they have been cast in a form similar to 
that of the Skyrme interaction. They had been already employed, under the form of a Yukawa 
function, in Ref.~\cite{suzuki} and they have been shown to reproduce well the correct 
mass number dependence of the Coulomb displacement energies, as well as 
a number of values of isospin mixing in the ground state. Finally, they turned out to be important 
to account for the IAR width in $^{208}$Pb ~\cite{colo98}. For all these reasons, we use these 
parametrizations in the present work. Because of their 
operatorial form, they do not add any contribution to the {\itshape p-h} force in 
RPA or QRPA. The electromagnetic spin-orbit is quite small: consequently, 
the associated energy shift has been added
to the HF-BCS results using first-order perturbation theory. 

\section{RESULTS}

The systematic trend of the IAR energies in the nuclei we have considered, $^{104\mbox{--}132}$Sn, 
is plotted in Fig.\ \ref{syst_trend}. The energies are obtained within QRPA, by including all the terms
mentioned in the previous Section: only the proton-rich 
$^{104,106}$Sn have been calculated using quasiparticle Tamm-Dancoff approximation (QTDA) 
because of QRPA instabilities.
Our findings are compared with the experimental energies quoted in 
Ref.~\cite{pham95}, where the results of the ($^3$He,$t$) reaction performed at  
an incident beam energy of 200 MeV are reported. It can be immediately 
realized that the agreement is fairly good. The difference between theory and experiment
is typically $\approx$ 200 keV in the series of isotopes which have been measured,
namely $^{112\mbox{--}124}$Sn (with the exception of the two extremes $^{112}$Sn and $^{124}$Sn
where this difference is larger). 
It is remarkable that another microscopic, self-consistent model like RMF --- which starts
from a quite different description of the nuclear mean-field and its oscillations as 
already stressed in the Introduction --- produces a similar numerical outcome~\cite{paar0304}. 
The results for the IAR in the unstable nuclei do constitute a useful guideline
for possible future experiments. 

Concerning the results in the ($N+1$,$Z-1$) channel, unfortunately 
few experimental measurements are available
for a comparision with our model. The only exception is the case of $^{120}$Sn. 
In Fig.\ \ref{In_channel} we plot for this nucleus the response to the IVGMR operator,
\begin{equation}
{\hat O}_{{\mbox {\upshape IVGMR}}} \equiv \sum_{i=1}^A r^2_i t_+(i),
\label{op_ivgmr}
\end{equation}
as a function of the energy with respect to the ground state of $^{120}$In. 
The continuous curve has been obtained by averaging the QRPA discrete strength 
distribution with a 1 MeV width Lorentzian curve.
We can compare our results with three experiments carried out by means of
different nuclear reactions. By using ($\pi^-,\pi^0$) at 165 
MeV~\cite{erell}, ($^{13}$C, $^{13}$N) at 50 MeV/A~\cite{berat} and 
($^{7}$Li, $^{7}$Be) at 350 MeV~\cite{annakkage} it has been shown, more or less
ambiguously, that a 0$^+$ state should lie, respectively, at 16.0$\pm$2.2 MeV, 14.7$\pm$1 MeV and
17.0$\pm$1.6 MeV. In our calculation most of the strength is found indeed
in the energy region between 12 and 20 MeV. Our main peak seems 
compatible with the ($^{7}$Li, $^{7}$Be) result. 

Coming back to the case of the IAR, we analyze in more detail our results
in order to clarify the most important features of our theoretical description. 
Firstly, in analogy with the conclusion drawn in Ref.~\cite{paar0304}, we may show that
also in the present case the consistent treatment of pairing correlations is
very important. In Fig.\ \ref{pn} we display three different results obtained
for the IAR strength distribution in $^{114}$Sn. Not only the residual proton-neutron
pairing force plays a crucial role to concentrate the IAR in a single peak; it 
also affects the IAR energy in an important way, that is, it induces a downward 
shift of about 500 keV. In the whole isotopic 
series we have studied, the peak associated with the IAR exhausts typically 
a percentage between 95\% and 98\% of the $N-Z$ sum rule. 
Only in the isotope $^{108}$Sn the IAR is found to be split in 
two peaks. 

Having assessed the importance of the proton-neutron residual pairing, we have also
tried to understand the role played by various other correlations present in
our model. For this purpose, we display in Fig.\ \ref{contrib} results for the
IAR energy in $^{120}$Sn obtained by making different approximations. 
The first number on the left side refers to a simple TDA calculation, without any 
pairing, without the spin-orbit term in the residual {\itshape p-h} force, and without
the other terms which have been often neglected (electromagnetic
spin-orbit, CSB and CIB). This would be the simplest possible calculation, 
analogous to that performed for many closed-shell nuclei in the previous literature.
The inclusion of RPA ground state correlations do not affect very much the
IAR, as it is expected for a nucleus which has neutron excess; the effect is
larger if we move towards the neutron-deficient isotopes. Pairing correlations
are more important. We have discussed above that they have to be included 
consistently (we repeat that a calculation with pairing only in the ground state would lead
to a too high, and fragmented, IAR): moreover, their inclusion shifts the IAR
downwards by about 150 keV. At this stage, the QRPA result would differ from
the experimental finding by about 500 keV. This would be approximately true for all the
stable isotopes. The two-body spin-orbit have a 
non negligible effect (about 100 keV) in pushing the IAR energy towards
the experimental value. Even more important, from a quantitative point of view, are in this case
the CSB and CIB forces which are inserted in the HF-BCS calculation (the
fact that they have opposite sign has been already remarked~\cite{auerbach_report}). 
Finally, we have included for the sake of
completeness the one-body electromagnetic spin-orbit. This term has also been
calculated long time ago (see, e.g, p. 494 of Ref.~\cite{nolen}) and 
it is known to have, as a rule, an effect of only few tenths of keV on the Coulomb 
displacement energies. Because of its $j$-dependence, it may become 
significant in the case of pure transitions associated with large angular
momentum, as it has been stressed in~\cite{ekman04}. We should add that
we have checked that the contributions stemming from the CSB, CIB and the 
electromagnetic spin-orbit are almost constant over the isotopic chain. In this
sense, the numbers presented in Fig.~\ref{contrib} are considered as
typical. As far as the two-body spin-orbit is concerned, in the middle of
the chain the associated repulsive contribution is maximum; at the extremes of the
chain it becomes smaller or even attractive (for instance, in $^{132}$Sn we 
find an attractive contribution associated with the diagonal $h_{11/2}$ matrix
element). 

Since many Skyrme parametrizations are available in the market, we would like
to mention that our results are not very sensitive to the choice of a specific
set. In fact, we have seen that the IAR energy of $^{120}$Sn varies by
less than 100 keV if we calculate it either using the force SLy4, or 
SIII~\cite{beiner} or SGII~\cite{sgii}. We have also performed a
calculation using the recently introduced SkO' interaction~\cite{skop}, in view
of the possibility of testing it in the next future on the systematics of spin
states. In this case, the variation of the energy, with respect to the result
obtained by using SLy4, is somewhat larger~\cite{note_sc}. Also the 
effect of varying the pairing strength $V_0$ has been
considered, and we refer to Fig.\ \ref{pairing_effect} for the results obtained
in the case of $^{116}$Sn (i.e., the isotope in the middle of the 50--82 neutron shell). 
We can consider as satisfactory that variations of $V_0$ in the 
range $\approx$ 650--710 MeV~fm$^{3}$, which 
lead to sizeable ($\approx$20\%) variations of $\Delta$, do not seriously affect 
the energy of the IAR. We can quite generally conclude that
the choice of parameters, both of our {\itshape p-h} and {\itshape p-p} forces, 
do not seriously affect our main conclusions on IAR.

\section{CONCLUSIONS}

Very few examples of microscopic, fully self-consistent charge-exchange 
QRPA calculations exist (in contrast with the non charge-exchange case).
This has motivated the present work, in which we have developed the method
and analyzed some specific issues: the relation between the isospin invariance
and the self-consistency in the pairing channel, and the role of the  
usually neglected contributions in the mean-field. We have applied our scheme 
for the calculation of the IAR along the chain of the Sn isotopes. Only 
calculations based on RMF are available for this case. We find that our 
non-relativistic model can account quite well for the experimental results. 

We plan to extend our calculations, and make further analysis of the charge-exchange
states. This will be done for different multipolarities, both in the non spin-flip and
spin-flip sectors. It is hoped that the comparison with experimental data, and
with the outcome of other microscopic models, can be instrumental to fix
rather general problems. In fact, as stressed in our Introduction, 
many uncertainties plague the isovector channel of the effective {\itshape NN} 
interaction, and consequently the symmetry part of the nuclear equation of 
state. 

A possible improvement of our model consists in changing the description of
the nuclear ground state, which may be calculated within full HFB instead 
of HF-BCS. This could allow a better description in the case, for instance, 
of weakly bound systems. Another open problem is the consideration of the 
role played by the proton-neutron pairing. Literature 
reflects the existence of many different thoughts about this 
interesting issue; a full microscopic QRPA  
calculation in the case in which the particles do not have a definite 
charge state may probably be at present too demanding. Finally, 
we should mention that the extension beyond mean-field of theories 
like ours remains to be done.

\newpage

\begin{figure}
\caption{The values of the pairing gaps $\Delta$ in the Sn isotopes. The open squares
correspond to the empirical values, extracted by attributing to the isotope with $N$ neutrons
the value which results from the three-point formula centered in $N+1$. 
The black squares correspond to the theoretical results: in this case, the
values of the state-dependent HF-BCS pairing gaps $\Delta_i$ are averaged in an energy
interval centered at the neutron Fermi energy and having a width of $\pm$5 MeV.}
\label{average_delta}
\end{figure}

\begin{figure}
\caption{Systematic trend of the IAR energies in the stable and unstable Sn isotopes. The
theoretical results, displayed by means of black circles, are compared with the experimental
data (open squares) whenever these are available. It must be noticed that the energies
are referred to the daughter nuclei.}
\label{syst_trend}
\end{figure}

\begin{figure}
\caption{Strength function associated with the IVGMR operator (\ref{op_ivgmr}) in 
$^{120}$In. The discrete QRPA peaks have been smoothed by using a Lorentzian averaging
(the Lorentzian width is 1 MeV). See the text for a comparison with the available
experimental results.}
\label{In_channel}
\end{figure}

\begin{figure}
\caption{Importance of the residual proton-neutron {\itshape p-p} interaction for the collectivity
of the IAR. The left, central and right panels refer respectively to RPA, QRPA without
that term in the residual force, and complete QRPA. The result is analogous to the one
shown in Fig.\ 5 of Ref.~\cite{paar0304}.}
\label{pn}
\end{figure}

\begin{figure}
\caption{Result for the IAR energy in $^{120}$Sn obtained using different approximations.
The values labelled by $\Delta$ represent the energy shifts of the IAR (in keV) at each step.
See the text for a detailed discussion.}
\label{contrib}
\end{figure}

\begin{figure}
\caption{Effect of the overall pairing strength $V_0$ which defines the effective
force (\ref{ppforce}) on the pairing gap (upper panel) and the IAR energy (lower
panel) in $^{116}$Sn. The experimental values are marked by horizontal full lines,
whereas the vertical dashed line indicates the adopted value of $V_0$.}
\label{pairing_effect}
\end{figure}


\begin{thebibliography}{9}
\bibitem{note_sc} However, we must stress that this is the only 
calculation of our work which is characterized by some (arguably  
small) self-consistency breaking. In fact, the calculation of the ground 
state includes the spin-gradient terms (the so-called
terms in $J^2$, $\vec J$ being the spin density) while the corresponding {\itshape p-h} residual
force is omitted.
\end{thebibliography}
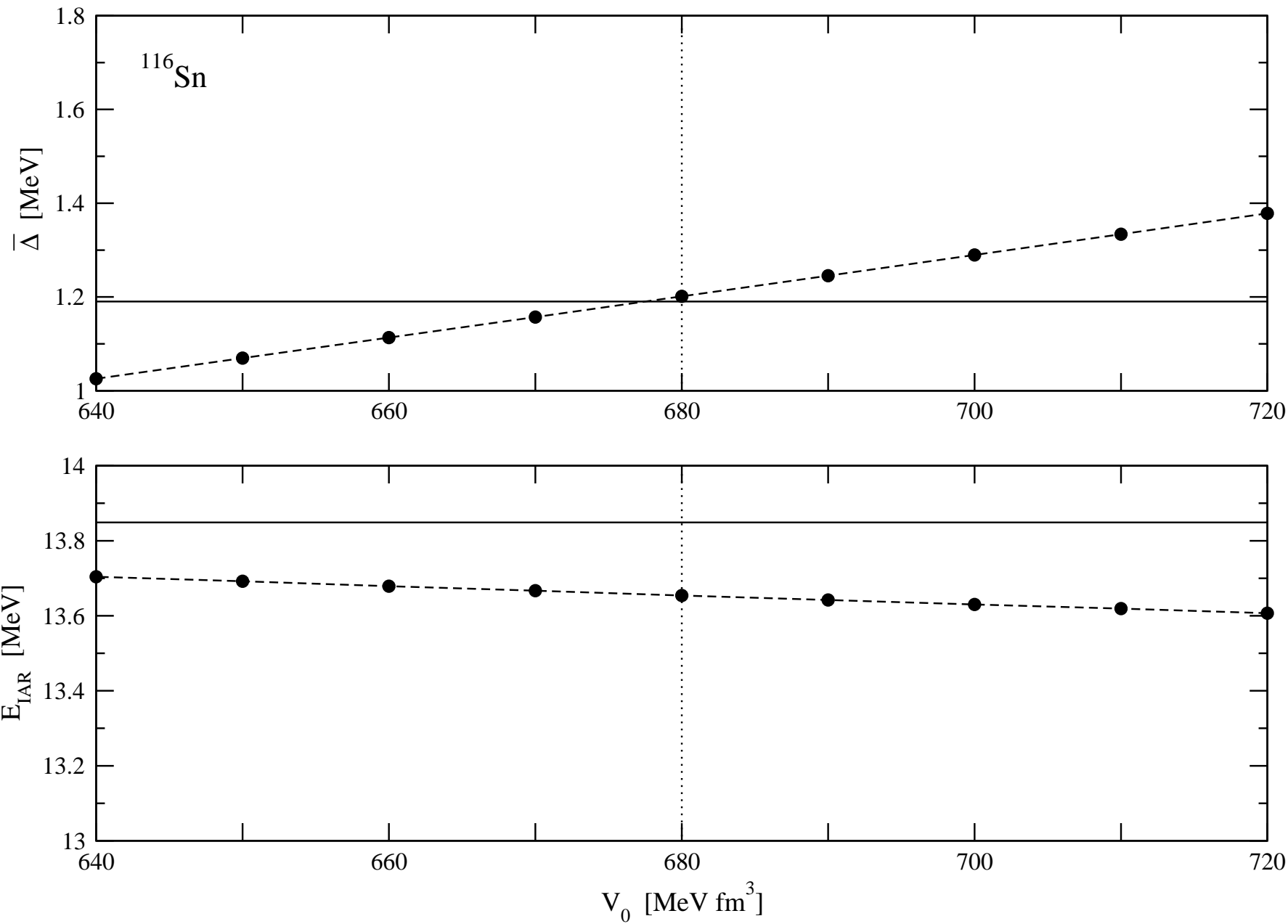
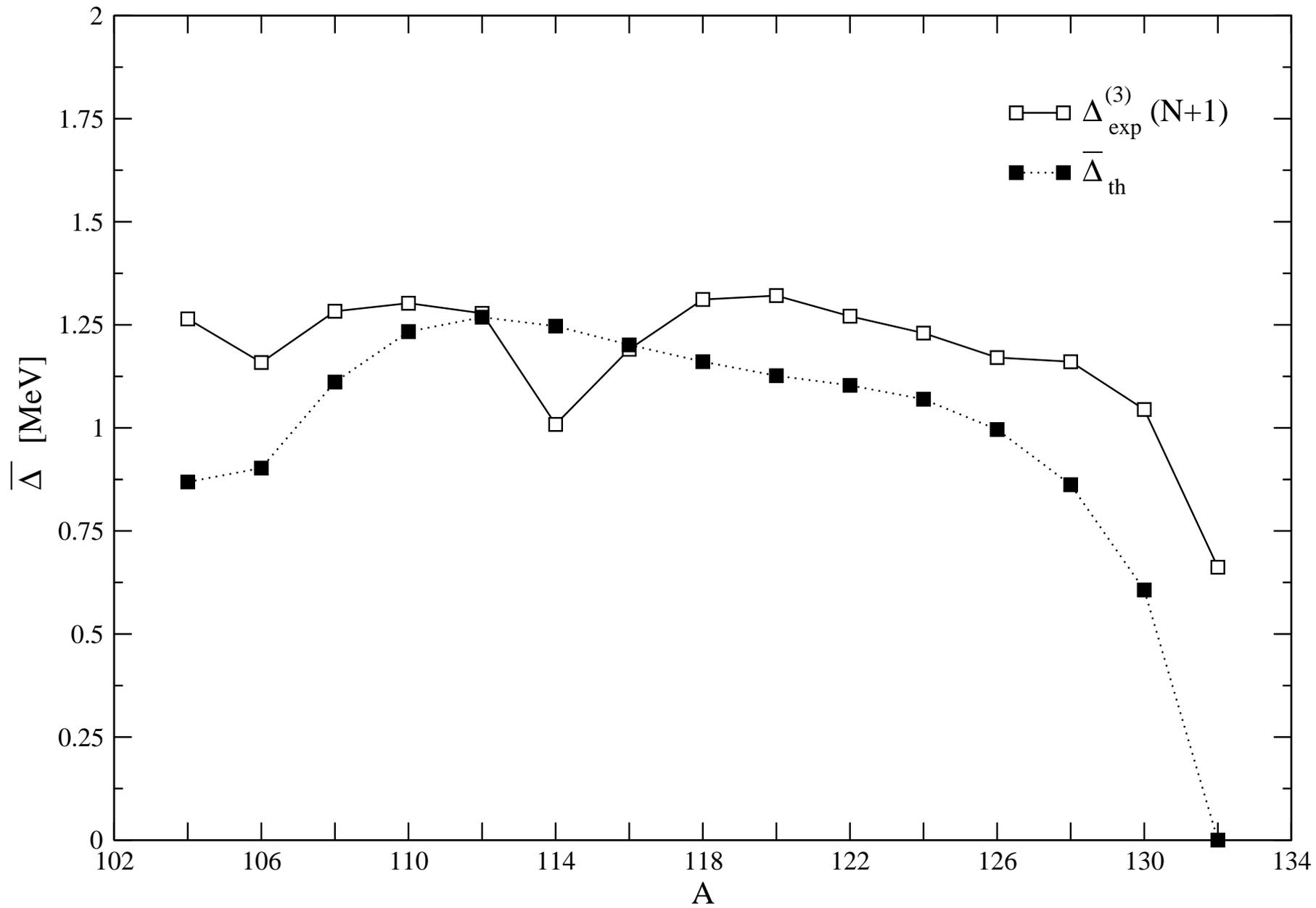
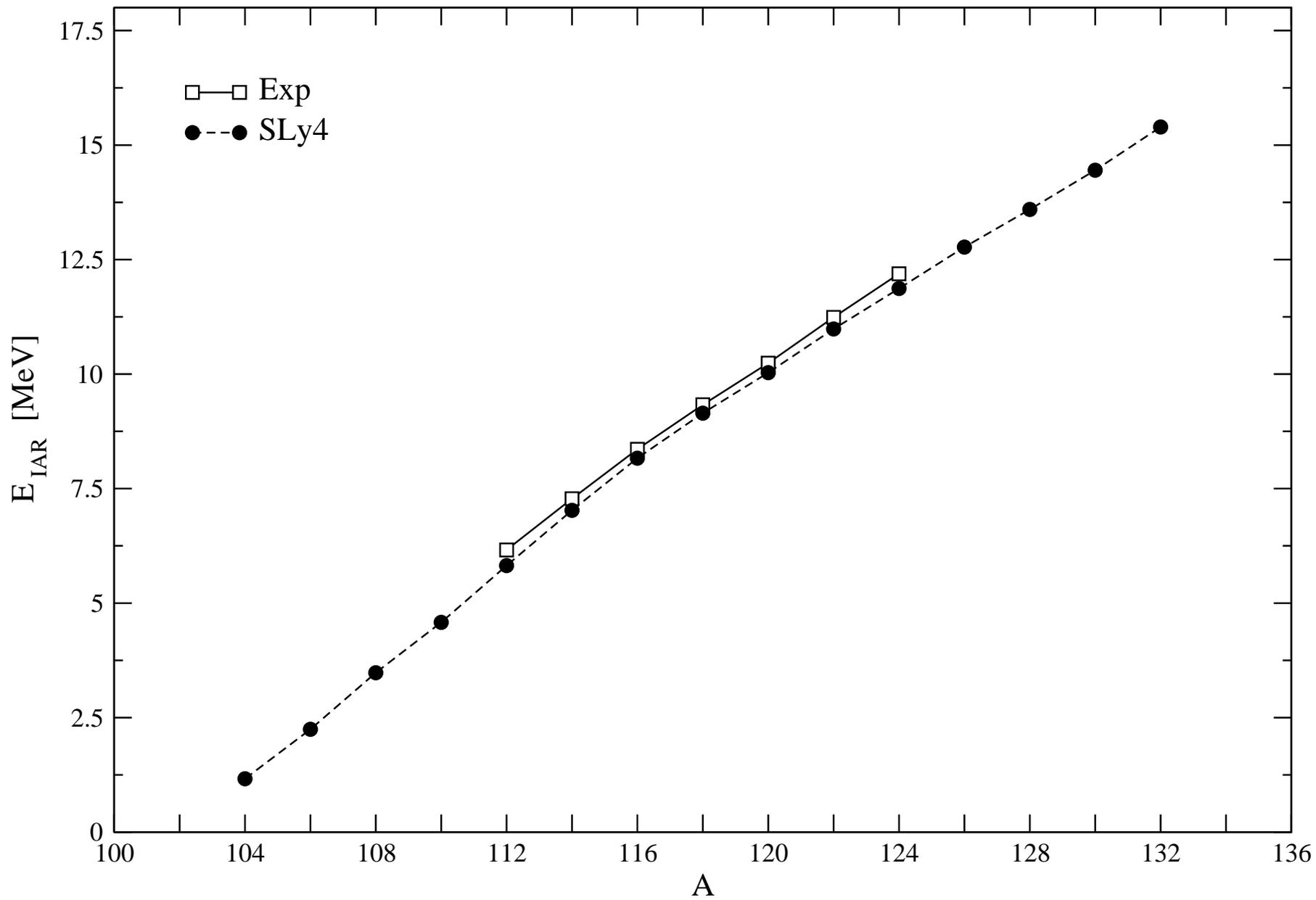
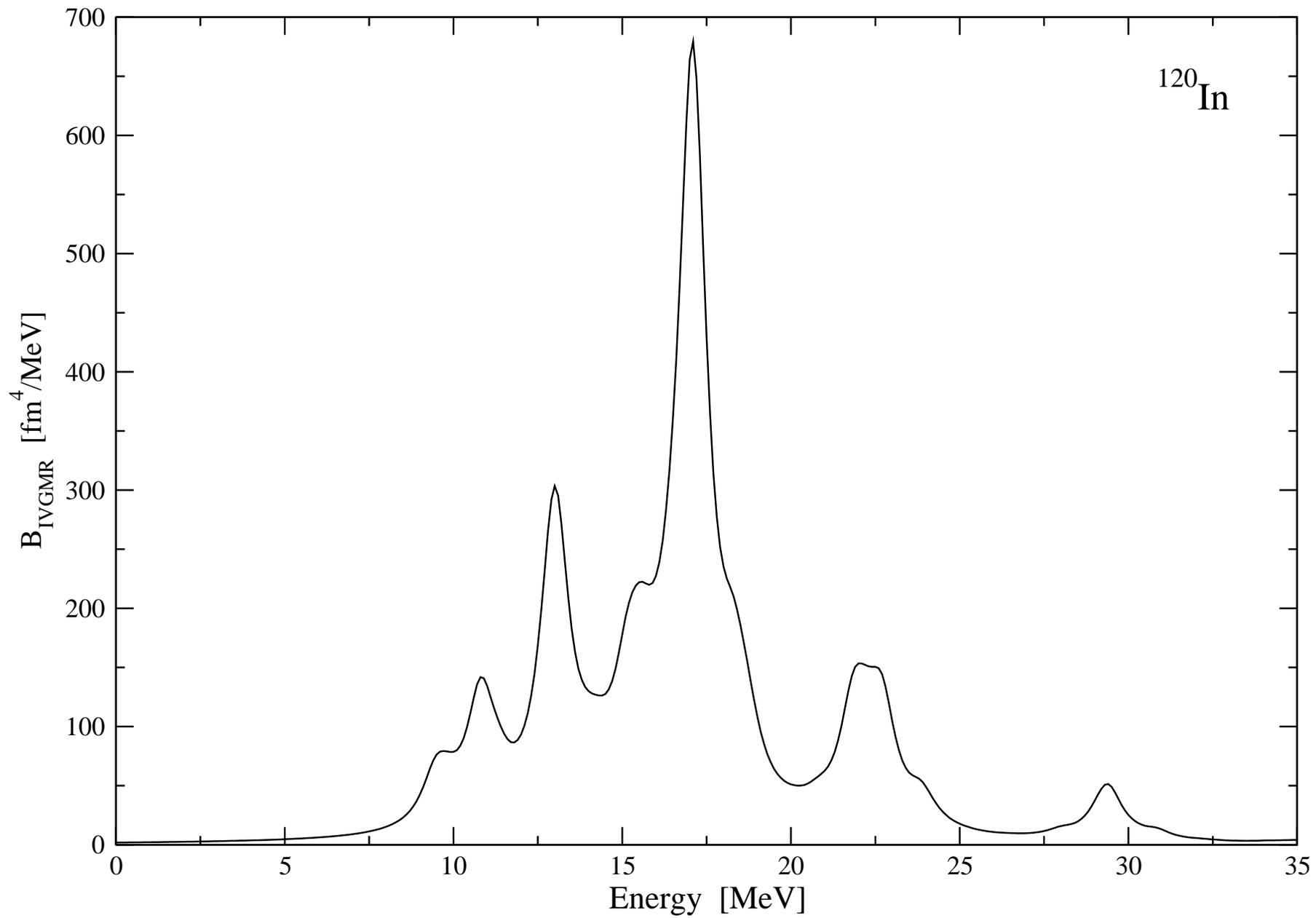
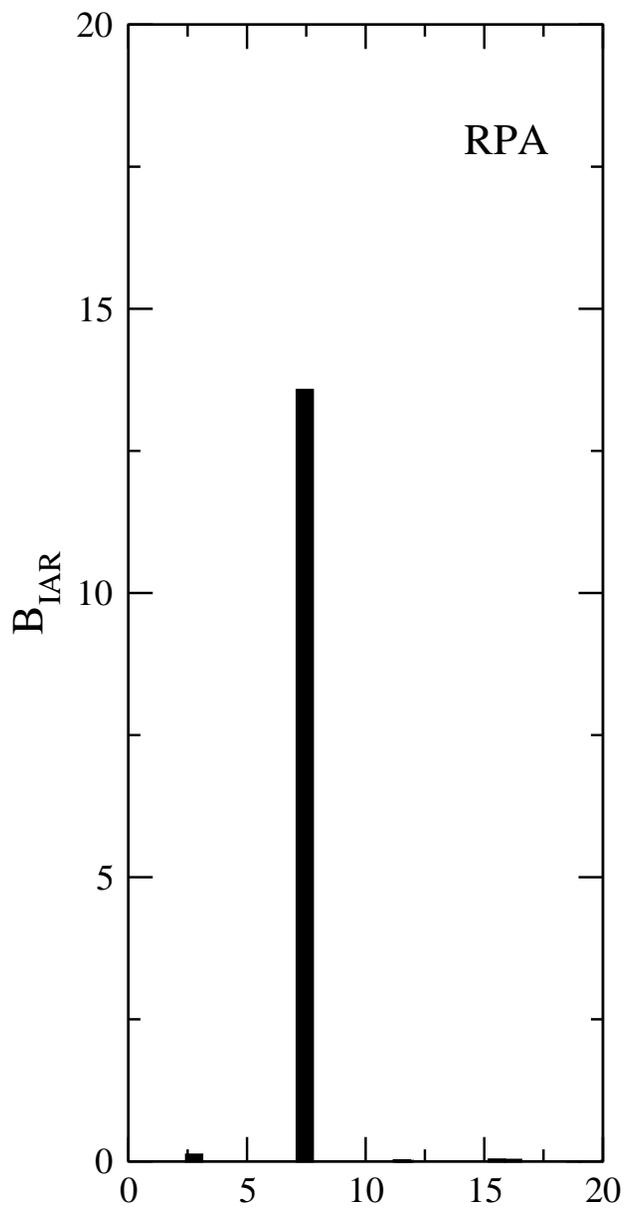
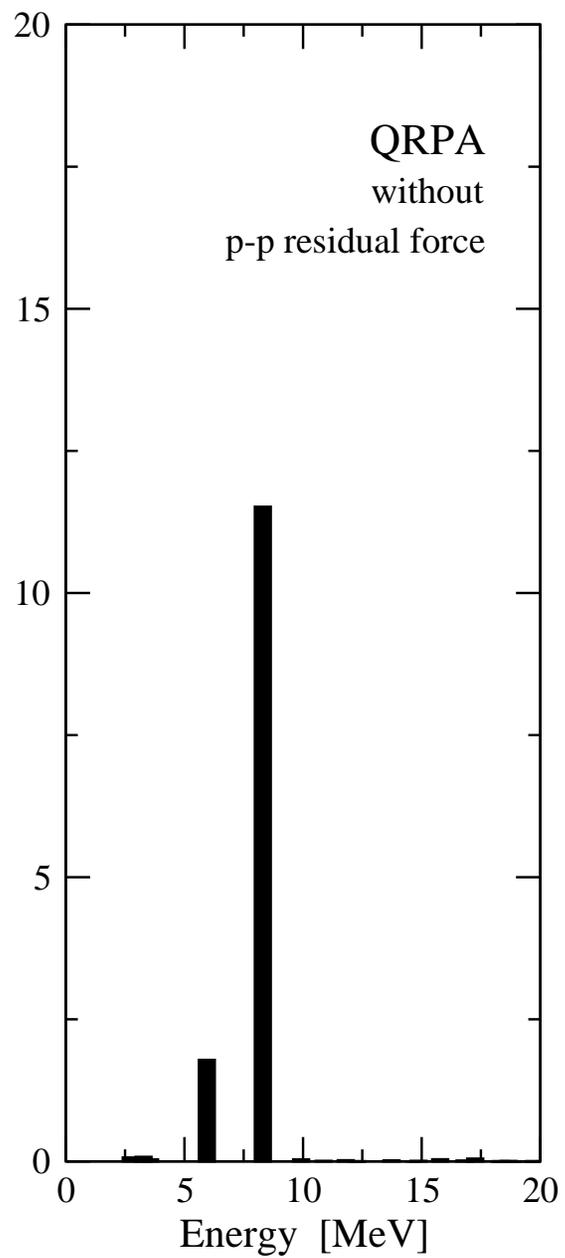
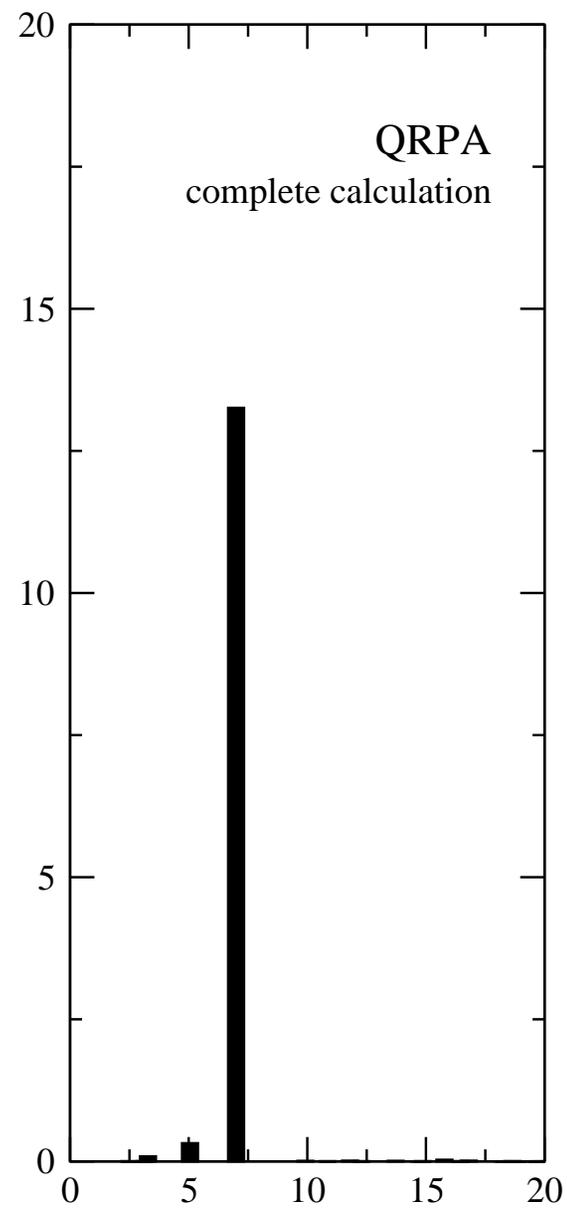
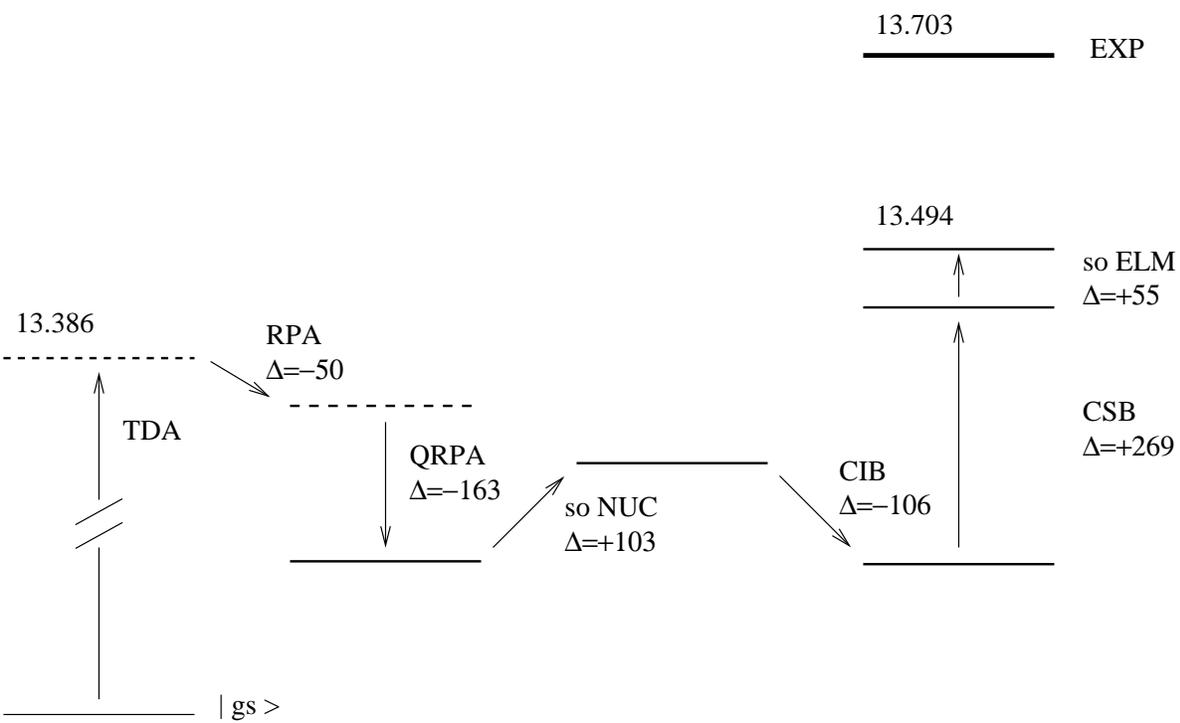
\end{document}